\documentclass{eptcs} %,noncommercial
\providecommand{\event}{FOCLASA 2015}
\usepackage[english]{babel}
\usepackage[T1]{fontenc}

\usepackage{latexsym}
\usepackage{hyperref}

\usepackage{listings}
\usepackage{mystyle}
\usepackage[usenames,dvipsnames]{xcolor}

\usepackage{cancel}

\allowdisplaybreaks 

\definecolor{bluekeywords}{rgb}{0.13,0.13,1}
\definecolor{greencomments}{rgb}{0,0.5,0}
\definecolor{turqusnumbers}{rgb}{0.17,0.57,0.69}
\definecolor{redstrings}{rgb}{0.5,0,0}

\lstdefinelanguage{FSharp}
                {morekeywords={let, new, match, with, rec, open, module, namespace, type, of, member, and, for, in, do, begin, end, fun, function, try, mutable, if, then, else, elif, when, null, %DA LEVARE: 
               dlet,tell,retract },
    keywordstyle=\bfseries\color{bluekeywords},
    sensitive=false,
    morecomment=[l][\color{greencomments}]{///},
    morecomment=[l][\color{greencomments}]{//},
    morecomment=[s][\color{greencomments}]{{(*}{*)}},
    morestring=[b]",
    stringstyle=\color{redstrings}
    }

\let\origthelstnumber\thelstnumber
\makeatletter
\newcommand*\Suppressnumber{%
  \lst@AddToHook{OnNewLine}{%
    \let\thelstnumber\relax%
     \advance\c@lstnumber-\@ne\relax%
    }%
}

\newcommand*\Reactivatenumber{%
  \lst@AddToHook{OnNewLine}{%
   \let\thelstnumber\origthelstnumber%
   \advance\c@lstnumber\@ne\relax}%
}

\makeatother

\begin{document} 

\title{A Context-Oriented Extension of F\# \thanks{Work partially supported by the MIUR-PRIN project Security Horizon.}}

%\numberofauthors{3}

\author{Andrea Canciani\qquad Pierpaolo Degano \qquad Gian-Luigi Ferrari \qquad Letterio Galletta 
\institute{Dipartimento di Informatica, Universit\`{a} di Pisa, Italy}\email{\{canciani,degano,giangi,galletta\}@di.unipi.it} }

\def\titlerunning{A Context-Oriented Extension of F\#}
\def\authorrunning{A. Canciani, P. Degano, G.-L. Ferrari \& L. Galletta}

\maketitle
 
\begin{abstract}
Context-Oriented programming languages provide us with primitive constructs to adapt program behaviour depending on the evolution of their operational environment, namely the context.
In previous work we proposed \coda, a context-oriented language with two-components: a declarative constituent for programming the context and a functional one for computing.
This paper describes the implementation of \coda\ as an extension of F\#.
\end{abstract}

\section{Introduction}
  % !TEX root = ../main.tex

Modern software systems are designed to operate \emph{always} and \emph{everywhere}, in partially known, ever changing operational environments. %, namely \emph{contexts}
Their structure is therefore subject to continuous changes that are unpredictable when applications are designed.
A suitable management of these changes should maintain the correct behaviour of applications and  their non-functional properties, e.g.\ quality of service.
Effective mechanisms are thus required to \emph{adapt} software to changes of the operational environment, namely the \emph{context} in which the application runs.

Using standard programming languages, context-awareness is usually implemented by modelling the context through a special data structure, which can answer to a fixed number of queries, whose results can be tested through \lstinline+if+ statements.
This approach however charges the programmer with the responsibility of implementing this data structure and the corresponding operations. % for the queries
In addition, she is responsible of achieving a good modularisation, as well as a good separation of cross-cutting concerns, and of the interactions between the code using the context and legacy code.
Successful examples are offered by special design patterns~\cite{Ramirez:2010} and Aspect-Oriented Programming~\cite{KiczalesHHKPG01} that encapsulate context-dependent behaviour into separate modules (objects and aspects, respectively), but leave most of the burden of handling the context and its changes (and their correctness) to the programmers.
Therefore, new dynamic development models and programming language constructs are required to effectively support the \emph{evolution} and \emph{adaptation} of applications to changes of the context~\cite{Baresi06}.

Recently, Context Oriented Programming (COP)~\cite{CostanzaH05} was proposed as a viable paradigm to develop context-aware software.
It advocates languages with suitable constructs for adaptation, so to express context-dependent behaviour in a modular manner.
In this way adaptivity is built-in, and the provided linguistic abstractions impose a good practice to programmers, with a positive effect on correctness and modularity of code, mainly because low level details about context management are masked by the compiler.

In previous papers~\cite{cop14,SEFM14} we followed this line of research from a foundational perspective and we proposed a programming language core, called \coda, specifically designed for adaptation and equipped with a clear formal semantics. 
It has two components: a logic constituent for programming the context and a functional one for computing.
%The \coda\ programming language proposes a new paradigm for developing adaptive systems. 
The logical component provides high level primitives for describing and interacting with complex working environments. 
The functional component supports programming of a variety of adaptation patterns. 
Its higher-order facilities are essential to exchange the bundle of functionalities required to manage adaptivity changes. 
Moreover, \coda\ offers a further support through a static analysis that guarantees programs to always be able to adapt in every context~\cite{SEFM14}.

In addition to the formal aspects of \coda\ studied in~\cite{SEFM14}, a main feature of our approach is that a single and fairly small set of constructs is sufficient enough for becoming a \emph{practical} programming language, as shown in the present paper.
Indeed, \coda\ can easily be embedded in a real programming eco-system, in our case .NET,  so preserving compatibility with its future extensions and with legacy code.
Being part of a well supported programming framework, our proposal minimises the learning cost and lowers the complexity of deploying and maintaining applications.
 
This paper illustrates a prototypical implementation of \coda\ as an extension of the (ML family) functional language F\# through a simple case study.
Indeed, no modifications at all was needed to the available compiler and to its runtime.
We exploited F\# metaprogramming facilities, such as code introspection, quotation and reflection, as well as all the features provided by .NET, including a vast collection of libraries and modules, and in particular a Just-In-Time (JIT) mechanism for compiling to native code.
All in all, \coda\ is implemented as a standard .NET library.\footnote{Available at \url{https://github.com/vslab/fscoda}}
In the path  towards the implementation a main role has been played by the formal description of the language.
In particular, the \coda\ formal semantics highlights and explains how the interactions between the two components occur. 
Indeed, the crucial parts of the implementation toolchain, compilation, generated code and run-time structures, together with their interactions, are \emph{formally} identified and described. 
As a matter of fact, the formal semantics gave us a great support, because it has been used to drive the definition of all the fine-grained control mechanisms governing \coda\ implementation.

\paragraph*{Plan of the paper}
In Section 2, we briefly review the design of the language and of its two components.
Section 3 is a gentle introduction to our COP language, through which we describe a simple case study, simulating an e-Healthcare system.
In Section 4, we present our approach to the implementation of  \coda\, that is then discussed in Section 5; the same section also briefly  compares our work with the relevant literature.
Finally, we draw some conclusions and present future work in Section 6. %surveys and

\section{Design of \coda}
  \label{sec:design}
  % !TEX root = ../main.tex

We survey \coda, a core functional programming language, equipped with linguistic primitives for context-awareness and coupled with a logical language, for context definition and management.
We first present how the context is described, and then the high level primitives for adaptation.
We omit discussing the two-phase static analysis designed for \coda~\cite{SEFM14}.
Briefly, it detects if an application will be able to adapt to its execution contexts:
at compile time, the first phase safely approximates the actions performed on the context;
at loading time the second phase exploits the approximation above to check that the program will adapt to \emph{all} the 
contexts that may occur at runtime.

\paragraph*{The context}

The notion of \emph{context}, i.e. the environment where applications run, is fundamental for adaptive software.
Intuitively, it is a heterogeneous collection of data coming from different sources and having different representations. 
Some of these data are application independent, like those about the hardware capabilities, e.g.\ screen resolution, and about the physical environment, e.g.\ the location of the user; other data are application-dependent like user's preferences, e.g.\ the language of the user.
It is worth noting that the context influences the shape and the features of the program input (e.g. from where it is taken) and how it is processed, but the context is \emph{not} part of the input itself. 
Of course, the context also affects the execution of the application.

In \coda, the context is split in two coarse parts: the \emph{system} and the \emph{application context}, following a well-established practice that separates data coming from inside the application and those coming from outside~\cite{Salehie:2009}.
Both parts are represented and manipulated in a uniform way, so guaranteeing compatibility and modularity.
Also, a programmer needs tools to access and manipulate all kind of contextual data in a easy and uniform way.
Indeed, the context is developed by requirements engineers~\cite{Zave:1997}, that have tools and skills different from those needed for building applications~\cite{Tanca2010,Loke:2004}.
This methodological issue, as well as separation of concerns motivated us to define \coda\ as a two-component language: a declarative constituent for programming the context and a functional one for computing.

The declarative approach allows programmers to express \emph{what} information the context has to include, leaving to the virtual machine \emph{how} this information is actually collected and managed.
For us, a context is a knowledge base and we  implement it as a Datalog program, following a well-studied approach~\cite{Tanca2010,Loke:2004}, within established methodologies~\cite{Kamina:2014b,Desmet:2007}.
In other words, a context in \coda\ is a set of facts that predicate over a possibly rich data domain, and a set of logical rules that permit to deduce further implicit properties of the context itself.
With this representation, adaptive programs can query the context and retrieve relevant information, by simply verifying whether a given property holds in it, i.e.\ by checking a Datalog goal.
Note that deduction in Datalog is fully decidable in polynomial time~\cite{CeriDatalog}.

\paragraph*{Adaptation}
As for programming adaptations, the functional part of the language provides two mechanisms.
The first is \emph{context-dependent binding} through which a programmer declares variables whose values depend on the context.
For that, we introduce the \lstinline+dlet+ construct that is syntactically similar to the standard \lstinline+let+, but has an additional Datalog goal therein:
\lstinline+dlet x = e1 when goal in e2+.
The variable \lstinline+x+ (called \emph{parameter} hereafter) may denote different objects, with different behaviour depending on the different properties of the current context, as required by \lstinline+goal+.
This is a major aspect of adaptivity.

The second mechanism is based on \emph{behavioural variations}: a chunk of behaviour that can be activated depending on information picked up from the context, so to dynamically adapt the running application.
This construct implements the fundamental concept of the COP paradigm, and in \coda\ it is a list of pairs \lstinline+goal.expression+, similar to pattern-matching, that alters the control flow of applications depending on the context.
When a behavioural variation is executed, parts of the deployed code are suitably selected
to meet the new requirements.
Behavioural variations have parameters and they are values, so they can be referred to by identifiers; passed and returned by functions; supplied by the context; and composed with existing ones.
This facilitates programming dynamic adaptation patterns, as well as reusable and modular code. 

Besides the features that describe and query the context, and those that adapt program behaviour, \coda\ is also equipped with constructs that update the context by adding data represented as Datalog facts \lstinline+F+ (\lstinline+tell F+) and removing them (\lstinline+retract F+).

\section{A simulator of an e-Healthcare system}
\label{sec:example}
  % !TEX root = ../main.tex

In this section we illustrate and discuss the main features of \coda\ and how these affect the development of an application. 
In the following example, we consider a fragment of an e-Healthcare system with a few aspects typical of the Internet of Things.
In particular, we discuss how physicians can access the medical data through their devices and how this system can help them to plan some medical activities.

\paragraph{An e-Healthcare scenario}
In our scenario each physician is provided with a device, e.g.\ a smartphone or a tablet, which tracks her location and 
enables her to retrieve a patient's clinical record.
On the basis of the obtained data, the doctor can decide which exams the patient needs and the system helps  scheduling them. 
Additionally, the system checks whether the doctor has the competence and the permission to actually perform the required exam, otherwise it may suggest another physician who can, possibly coming from another department. 
When a doctor moves from a ward to another, her operating context changes, in particular she can access the complete clinical records of the patients therein. 
The application must adapt to the new context and it may additionally provide different features, e.g.\ by disabling rights to use some equipment and by acquiring access to new ones. 

\paragraph{The e-Healthcare context}
\lstset{language=Prolog,mathescape,basicstyle=\ttfamily\small,showstringspaces=false}
We consider a small portion of our e-Healthcare system, and we show how to declaratively describe the relevant contextual information, through Datalog.
In particular, we take into account the part of the context that stores and makes available some data about the doctors' location, information about their devices, the patients' medical records and the ward medical equipment. 
Some basic data are asserted by Datalog facts, and one can retrieve further information through the inference machinery of Datalog, that uses logical rules, also stored in the context.

For example, the fact that Dr.\ Turk is in cardiology is rendered by the fact
\begin{lstlisting}
physician_location('Dr. Turk', 'Cardiology').
\end{lstlisting}

The following inference rule permits to deduce that a doctor can access the clinical data of patients in the same department where she is:
\begin{lstlisting}
physician_can_view_patient(Physician, Patient) :-
        physician_location(Physician, Location),
        patient_location(Patient, Location).
\end{lstlisting}
This rule states that the predicate on the left hand-side of the implication operator \lstinline+:-+ holds when the conjunction of the predicates (\lstinline{physician_location} and \lstinline{patient_location}) in the right hand-side yields true, i.e.\ when the physician and patient's location are the same.
The \coda\ context is quite expressive and allows us to model fairly complex situations. 
For example, sometimes the patients could be
prescribed an exam which can only be performed after some other screenings.
Therefore, to compute the list of exams a patient needs, we have to take into account all the dependencies among them.
This could be encoded in the context through the following recursive rules:

\begin{lstlisting}
patient_needs_result(Patient, Exam) :-
        patient_has_been_prescribed(Patient, Exam).
  
patient_needs_result(Patient, Exam) :-
        exam_requirement(TargetExam, Exam),
        patient_needs_result(Patient, TargetExam).
\end{lstlisting}

The first rule states that the prescription of an exam implies that the involved patient needs the results of the test.
The second rule says that whenever a patient needs an exam, she also needs all the screenings the exam depends on. 
Datalog provides a convenient way to model recursive relations like the dependency among exams, that may require involved queries with standard relational databases.

The next rule dictates that a patient has to do an exam if the two clauses in the right hand-side are true.
The first has been already discussed above, while
the second clause says that a patient should \emph{not} do an exam if its results are already known (in the rule below the operator \lstinline!\+! denotes the logical \emph{not}). \footnote{ %
Our version of Datalog only admits safe and stratified programs, so to effectively cope with negation~\cite{CeriDatalog}.}

\begin{lstlisting}
patient_should_do(Patient, Exam) :-
        patient_needs_result(Patient, Exam),
        \+ patient_has_result(Patient, Exam).
\end{lstlisting}

In addition, physical objects can be declaratively described in quite a similar, homogeneous manner.
The following (simplified) rule specifies when a device can display a certain exam, by checking whether it has the needed capabilities:

\begin{lstlisting}
device_can_display_exam(Device, Exam) :-
        device_has_caps(Device, Capability),
        exam_view_caps(Exam, Capability).
\end{lstlisting}
   
Asserting the capabilities of a device is straightforward, by listing a set of facts, e.g.

\begin{lstlisting}
device_has_caps('iPhone 5', '3D acceleration').
device_has_caps('iPhone 5', 'Video codec').
device_has_caps('iPhone 5', 'Text display').
device_has_caps('Apple Watch', 'Text display').
\end{lstlisting}

\paragraph{Adaptation constructs}
\lstset{language=FSharp,basicstyle=\footnotesize\ttfamily,showstringspaces=false,showspaces=false}

Now, we show how context-dependent bindings and behavioural variations allow a programmer to express program behaviour which depends on the context in our e-Healthcare system. 
When a doctor enters a department and visits some patients, she can display the patients' medical records on her personal device.
Moreover, the e-Healthcare system computes the list of the clinical exams a patient should do and that the doctor can perform.
The following code implements these functionalities, and shows all the adaptation constructs of \coda.
The \lstinline+display+ function below takes a doctor \lstinline+phy+ and a patient \lstinline+pat+ as arguments and prints on the screen the information about the patient's exams.

\begin{lstlisting}[numbers=left]
let display phy pat =
  match ctx with
  | _ when !- physician_can_view_patient(phy, pat) ->
    match ctx with
    | _ when !- patient_has_result(pat, ctx?e) ->
      printfn "%s sees that %s has done:" phy pat
      for _ in !-- patient_has_result(pat, ctx?exam) do
        display_exam phy ctx?exam
    | _ ->
      printfn "%s sees that %s has done no exam" phy pat

    let next_exam = "no exam" |- True
    let next_exam = ctx?exam |-
                      (physician_exam(phy, ctx?exam),
		       patient_active_exam(pat, ctx?exam))
    printfn "%s can submit %s to %s" phy pat next_exam
  | _ ->
    printfn "%s cannot view details on %s" phy pat
\end{lstlisting}

Actually, the code above is F\#, a dialect of ML.
Since we wanted to keep the F\# parser unmodified, the syntax is slightly different from the one used in~\cite{SEFM14} and recalled in Section~\ref{sec:design}.
This is only an implementation detail, because a simple macro-expansion suffices to translate the original syntax in the intermediate notation used in this section.

%, i.e. \lstinline+goal.expression+. 
A behavioural variation has the form \lstinline+match ctx with | _ when !- Goal -> expression+.  
The sub-expression \lstinline+match ctx with+ explicitly refers to the context; the part \lstinline+| _ when !- Goal+ introduces the goal to solve; and \lstinline+-> expression+ is the sub-expression to evaluate when the goal is true.

The outermost behavioural variation (starting at line 2) checks whether the doctor \lstinline+phy+ is allowed to inspect the data of the patient \lstinline+pat+, as granted when the goal \lstinline+physician_can_view_patient(phy, pat)+ at line 3 holds.

The nested behavioural variation (line 4) checks if the patient has got the results of some exams, through the predicate \lstinline+patient_has_result+.
If this goal holds, the \lstinline+for+ construct extracts the list of exams and results from the context (line 7).
The statement \lstinline+for _ in !-- Goal do expression+  iterates the evaluation of \lstinline+expression+ over all the solutions of the \lstinline+Goal+.
In other words, it is an iterator on-the-fly, driven by the solvability of the goal in the context.
Note that the predicate \lstinline+patient_has_result+  at line 7 contains the \emph{goal variable} \lstinline+ctx?exam+: if the query succeeds, at each iteration \lstinline+ctx?exam+ is bound to the current value satisfying \lstinline+Goal+.
A goal variable is introduced in a goal, defining its scope, through the syntax \lstinline+ctx?var_name+.

Finally, the function \lstinline+display+ prints an exam that can be performed next on the patient \lstinline+pat+ by the physician \lstinline+phy+, by using the construct for the context-dependent binding.
Here we write it in the following form \lstinline+let x = expression1 |- Goal [in] expression2+, where the keyword \lstinline+let+ and the operator \lstinline+|-+ replace \lstinline+dlet+ and \lstinline+when+, used in  Section~\ref{sec:design}.
At lines 12-13 we declare by cases the parameter \lstinline+next_exam+, that is referred to in line 16.
Which case applies and which value will be bound to \lstinline+next_exam+ can only be determined at runtime when the parameter is used, because they depend on the actual context.
If the goal in lines 14-15 hold, then  \lstinline+next_exam+ assumes the value retrieved from the context, otherwise it gets the default value \lstinline+"no exam"+. 
(The predicate \lstinline+True+ always holds independently of the context.)

Actually, it may happen that no goal is satisfied in a context during the execution of a behavioural variation or during the resolution of a parameter.
This reflects the inability of the application to adapt, either because the programmer assumed at design time the presence of
functionalities that the current context lacks, or because of programming errors. 
This is a new class of runtime errors that we call \emph{adaptation failures}. 
For example, the following function assumes that given the identifier of a physician, it is always possible to retrieve her location from the context through the \lstinline+physician_location+ predicate:

\begin{lstlisting}
let find_physician phy =
  let loc = ctx?location |-
              physician_location(phy, ctx?location) in
  loc
\end{lstlisting}

If the function \lstinline+find_physician+ is invoked on a
physician whose location is not stored in the context, e.g.\ due to a programming error, the context-dependent binding will fail to
find a solution for the goal.
In such a case the current implementation throws a runtime
exception. 
This is a common pattern in languages like F\#, which makes
behaviour both easy to manage and to integrate with existing code. 
In the current implementation we use the standard F\# construct to handle an adaptation failure as shown in the following snippet of code:

\begin{lstlisting}
let find_physician phy =
  try
    let loc = ctx?location |-
                physician_location(phy, ctx?location) in
    loc
  with e ->
    printfn "WARNING: cannot locate %s:\n%A" phy e
    "unknown location"
\end{lstlisting}

A more sophisticated approach involves statically determining whether
the adaptation might fail and reporting it before running the
application, as described in \cite{SEFM14}.

The interaction with the Datalog context is not limited to queries: it
is possible to program the modifications to the knowledge base on which it
performs deduction. 
Adding or removing facts is done by the \lstinline+tell+ and \lstinline+retract+ operations, as in:

\begin{lstlisting}
tell <| patient_has_result("Alice", "CT scan")
\end{lstlisting}

\paragraph{Some execution examples}

We now show how the functions defined above give different results when invoked in different contexts, some parts of which will only be described intuitively.\footnote{ %
For a full definition of the code see  \url{https://github.com/vslab/fscoda}.}

For example, in a context where Dr. Turk is not in the same ward as Bob

\begin{lstlisting}
display "Dr. Turk" "Bob"
\end{lstlisting}

\noindent outputs
\begin{verbatim}
Dr. Turk cannot view details on Bob
\end{verbatim}

\noindent
because physicians are only allowed to see data about the
patients in the department where they are. 
In particular, the behavioural variation on
\lstinline+physician_can_view_patient+ at line 3 finds out that the operation is not
allowed. 
If instead Dr.\ Cox is in the same department where Bob is, the call

\begin{lstlisting}
display "Dr. Cox" "Bob"
\end{lstlisting}

\noindent
correctly prints the details about Bob (actually these are stored in the Datalog knowledge base):

\begin{verbatim}
Dr. Cox sees that Bob has done no exam
Dr. Cox can submit Bob to Blood test
\end{verbatim}

In this case the outermost behavioural variation (starting at line 2) confirms that Dr. Cox can view the data. 
The nested one (starting at line 4), driven by \lstinline+patient_has_result+, finds no exam for Bob, hence the
function displays the no-exam message (line 10).
Moreover, the program finds out that Dr. Cox could do a blood test on Bob, as he is enabled to, and additionally
Bob needs no pre-screening and so that exam can be done immediately, because the predicate at line 15 holds.

We now consider a slightly more complex situation, in which the context itself is modified.
Patient Alice has already performed an EEG test, and doctors prescribed her a CT and nothing else.
Dr. Kelso is in Alice's room, is enabled to do only CT tests and carries a device on which he can visualise the results.
In this context

\begin{lstlisting}
display "Dr. Kelso" "Alice"
\end{lstlisting}

\noindent
outputs

\begin{verbatim}
Dr. Kelso sees that Alice has done:
  - EEG
Dr. Kelso can submit Alice to CT scan
\end{verbatim}

The main difference from the case above is that Alice has already performed
an exam, which is hence listed by the iteration construct. 
Now Dr. Kelso performs a CT scan on Alice and thus the context has
to be accordingly changed, by asserting the fact

\begin{lstlisting}
tell <| patient_has_result("Alice", "CT scan")
\end{lstlisting}

The same fragment of code above 
\begin{lstlisting}
display "Dr. Kelso" "Alice"
\end{lstlisting}
\noindent
has a different output in the modified context
\begin{verbatim}
Dr. Kelso sees that Alice has done:
 - EEG
 - CT scan
Dr. Kelso can submit Alice to no exam
\end{verbatim}

Besides displaying a longer list of exam results, the application shows Dr. Kelso that Alice needs him to
perform no other exam. 

Assume Dr. Cox moves to Alice's room and checks her medical report, but he has a device that cannot show CT images

\begin{lstlisting}
display "Dr. Cox" "Alice"
\end{lstlisting}

prints the following

\begin{verbatim}
Dr. Cox sees that Alice has done:
 - EEG
 - CT scan (current device cannot display the exam data)
Dr. Cox can submit Alice to no exam
\end{verbatim}

Since the CT scan cannot be displayed by the device, the \lstinline+display_exam+ function
warns the doctor and it might present the results in a more limited form, e.g\ a
static thumbnail.

The e-Healthcare system might also help Bob in finding out the physicians that can
visit him:

\begin{lstlisting}
for _ in !-- (patient_active_exam("Bob", ctx?exam),
	      physician_exam(ctx?physician, ctx?exam)) do
  printfn "%s (currently in %s) can submit Bob to %s"
    ctx?physician (find_physician ctx?physician) ctx?exam
\end{lstlisting}

This fragment prints the name and the position of all of the physicians
who can perform an exam which Bob needs. This code snippet relies on the second
version of \lstinline+find_physician+, which handles adaptation failures
by logging the error and returning \lstinline+"unknown location"+ as a
result. Assuming that Dr. Turk has left the hospital and Dr. Cox is in the cardiology department, the output would be:

\begin{verbatim}
Dr. Cox (currently in Cardiology) can submit Bob to Blood test
WARNING: cannot locate Dr. Turk:
CoDa.InconsistentContext: Context inconsistency detected
Dr. Turk (currently in unknown location) can submit Bob to Blood test
\end{verbatim}

Since it is not possible to deduce the location of Dr. Turk from the
context, the context-dependent binding in \lstinline+find_location+
fails. Nonetheless the program can continue the execution after
handling the exception.

\section{Internals of a prototype compiler}\label{sect:implementation}
  % !TEX root = ../main.tex

In order to implement \coda, we found it convenient to build upon a functional language and to integrate it with a Datalog engine, as easily as possible.
As said, our choice has been F\# which is commercially supported, and fully integrated in the .NET eco-system.
We exploited the metaprogramming facilities available in F\# for the \coda\ adaptation constructs to avoid re-implementing well-known primitives, while the deduction engine is that of YieldProlog, available as a .NET library.
The bipartite nature of \coda\ fosters the independence between context and application development.
This design choice is well supported by .NET through the notion of \emph{assemblies}. 
They work just as modules and offer us a natural way to separate the
 code of the context from that of the application.

\paragraph{Implementing the context}
 
A requirements engineer writes several Datalog sources describing the context in hand. 
These files are ahead-of-time translated to a .NET code by our compiler \verb+ypc+, which is based on a \emph{customised} version of the library YieldProlog.\footnote{Available at~\url{https://github.com/vslab/YieldProlog}} YieldProlog works as our Datalog engine, handling the context state and solving goals.
Our translation compiles each predicate into a method, whose code enumerates one by one the solutions, i.e.\ the assignments of values to variables which satisfy the predicate.
In this way, the interaction and the data exchange between the application and the context is fully transparent to the programmer because the .NET type system is uniformly used everywhere.
Indeed, data inside the context are instances of the class \verb+object+, hence the programmer can insert any object in the context as long as the method \verb+Equals+ is appropriately overridden. 
This is required because the Datalog engine during the deduction process needs to check if two objects (considered as atoms by the engine) are equal.

For example, the predicate
\lstset{language=Prolog,mathescape, basicstyle=\ttfamily\small}
\begin{lstlisting}
patient_needs_result(Patient, Exam) :-
       patient_has_been_prescribed(Patient, Exam).
  
patient_needs_result(Patient, Exam) :-
       exam_requirement(TargetExam, Exam),
       patient_needs_result(Patient, TargetExam).
\end{lstlisting}

\noindent
is translated into

\lstset{language=Java,basicstyle=\footnotesize\ttfamily,showstringspaces=false,showspaces=false} %breaklines=true,
\begin{lstlisting}[numbers=left]
public static IEnumerable<bool> patient_needs_result(object Patient, object Exam)
{
        foreach (bool l2 in patient_has_been_prescribed(Patient, Exam))
            yield return false;
            
        Variable TargetExam = new Variable();
        foreach (bool l2 in exam_requirement(TargetExam, Exam))
            foreach (bool l3 in patient_needs_result(Patient, TargetExam))
                yield return false;
}
\end{lstlisting}

The enumeration of goal solutions works through side-effects, by modifying the values of the input parameters, i.e.\ the return value (which is always \lstinline+false+) is never used.
The loop at lines 3-4 returns to the caller for each solution of \lstinline+patient_has_been_prescribed+, because the \lstinline+Patient+ and \lstinline+Exam+ pair is a solution according to the first rule.
At line 6, the compilation of the second rule introduces the variable \lstinline+TargetExam+, because it appears free in the body.
The conjunction of subgoals is obtained by two nested loops, so that the statement at line 9 is reached only if both loops enumerate consistent solutions.
The unification algorithm, implemented by the class \lstinline+Variable+, ensures the consistency of the solutions computed by the different subgoals.
The recursive definition of \lstinline+patient_needs_result+ requires no special handling as it is implemented as a recursive method in a straightforward way.
\paragraph{Functional part}

The application programmer writes F\# code,
annotating the functions which use \coda\ extensions with  \emph{custom attributes} (see code below) and starting the \coda\ runtime.
Since the operations needed to adapt the application to contexts are transparently handled by our runtime support, the  compiler \verb+fsharpc+ works as it is.
Indeed, the \coda-specific constructs are just-in-time replaced in a single step by their F\# implementation when they are about to be run. 
A simple example follows.
\lstset{language=FSharp,basicstyle=\footnotesize\ttfamily,showstringspaces=false,showspaces=false} %breaklines=true,

\begin{lstlisting}[numbers=left]
[<CoDa.Code>]
module Physicians.Test

open CoDa.Runtime       // the runtime
open Physicians.Facts   // functional to logic typed interface
open Physicians.Types   // logic to functional typed interface
open ...                // other libraries used by the application


[<CoDa.Context("pysician-ctx")>]
[<CoDa.Context("patients-ctx")>]
[<CoDa.Context("devices-ctx")>]
[<CoDa.EntryPoint>]
let main () =
  display "Dr. Turk" "Bob"
  display "Dr. Cox" "Bob"
  printfn ""
  display "Dr. Cox" "Alice"
  display "Dr. Kelso" "Alice"
  printfn ""

  // Other code
do
  run ()
\end{lstlisting}

The attribute \verb+CoDa.Code+ marks the \coda-specific
constructs which need to be transformed, e.g.\ the above  module \verb+Physicians.Test+.
Actually, the attribute  \verb+CoDa.Code+ is an alias for the standard \verb+ReflectedDefinitionAttribute+, that marks modules and members whose abstract syntax trees (AST) are used at runtime through reflection. 
Note that \coda-specific operations are only allowed in methods marked with this attribute; otherwise an exception is raised when they are invoked.

The attribute \verb+CoDa.EntryPoint+  marks the principal function of the application (\verb+main+ above).  
When the \coda\ runtime is initialised and started through the function \verb+run+, it looks for the function  \verb+f+ marked by \verb+CoDa.EntryPoint+; then it transforms the code of the function
replacing the \coda-specific constructs with their F\# object code; and finally runs the obtained object code.
The translation is performed on the AST represented in the form of
quotations. 
Hereafter, for readability, we will show the object code in F\# syntax, rather than the quotation emitted by the JIT compiler.

The lines 10-12 in the code show that the context is conveniently split in distinct modules. 
The attribute \verb+CoDa.Context+ describes which of its parts are needed for the application. 
The runtime initialises the context and links it with the application,  before running it.
The code that initialises the context of the e-Healthcare system behaves as the following one:
\lstset{language=FSharp,basicstyle=\footnotesize\ttfamily,showstringspaces=false,showspaces=false} %breaklines=true,

\begin{lstlisting}
[<CoDa.Code>]
module Physicians.Context

// code to import Runtime, Types, Facts (see lines 4-7 in the code above)

[<CoDa.ContextInit>]
let initFacts () =
  tell <| physician_exam("Dr. Cox", "ECG")
  tell <| physician_exam("Dr. Cox", "Blood test")
  // other code
\end{lstlisting}

The \coda\ construct \verb+tell+ adds facts in the context (there is also a \verb+retract+ to remove facts); it is implemented as a method of the \verb+Runtime.context+ object, only accessible by the \coda\ runtime.
%}
The modules \verb+Facts+ and \verb+Types+  provide the interface between the functional code and the context. 
In particular, they contain utility functions for typing facts and predicates in F\#, respectively.
The above function \lstinline+initFacts+ is compiled as
\begin{lstlisting}
let initFacts () =
  Runtime.context.Tell <| physician_exam("Dr. Cox", "ECG")
  Runtime.context.Tell <| physician_exam("Dr. Cox", "Blood test")
  // other code
\end{lstlisting}

We now discuss how the two main constructs for expressing adaptation, \emph{behavioural variations} and \emph{context-dependent
binding}, are implemented in our e-Healthcare system.
Consider the function \lstinline+display+ defined in Section~\ref{sec:example}.
The behavioural variations of lines 2-5 are compiled as
\begin{lstlisting}
if Runtime.context.Solve([|physician_can_view_patient(phy, pat)|], 
                         null) then
  let solution1 = new Dictionary<string, obj>()
  solution1.["e"] <- new Variable()
  if Runtime.context.Solve([|patient_has_done(pat, solution1.["e"])|], 
                           solution1) then
    ...
  else
    printfn "%s sees that %s has done no exam" phy pat
else
  printfn "%s cannot view details on %s" phy pat
\end{lstlisting}

The \lstinline+match+ expression becomes an \lstinline+if+, whose guard queries the context by the method \lstinline+Solve+. 
The above translation also illustrates the mechanism that implements the goal variables.
The dictionary \lstinline+solution1+ is initialised with the variable introduced in the goal, namely \lstinline+ctx?e+ of line 5. The dictionary is then passed to the Datalog solver as second argument.   
If a solution is found, \lstinline+solution1+  contains an assignment to the value that satisfies the goal, and allows us to access the solution, but only within the scope of the corresponding \lstinline+if+ expression.
Note that the outermost behavioural variation uses no goal variables, hence no dictionary is needed  (the second argument of \lstinline+Solve+ is \lstinline+null+). 

The \lstinline+for+ construct follows the same schema; the iteration of line 7 is translated as 

\begin{lstlisting}
let solution2 = new Dictionary<string, obj>()
solution2.["exam"] <- new Variable()
let variables0 = new Dictionary<string, obj>(solution2)
for _ in Runtime.context.Enumerate([|patient_has_done(pat, solution2.["exam"])|],
                                   variables0, solution2) do
  Runtime.callTramp display_exam [| phy; solution2.["exam"] |]
\end{lstlisting}
The method \lstinline+Enumerate+ returns a stream that iterates over the goal solutions by storing them in the dictionary \lstinline+solution2+. The original body of the \lstinline+for+ loop consisted of a call to the function \lstinline+display_exam+ (line 8), which is instrumented by interposing the method \lstinline+callTramp+. This is the hook used by the JIT compiler to overtake control when an \coda\ function needs to be translated. For merely technical reasons, the dictionary \lstinline+variables0+, which initially is a clone of \lstinline+solution2+, is passed to the method \lstinline+Enumerate+.

The binding of the parameter \lstinline+next_exam+ (lines 12-16) is implemented by the following snippet
\newpage
\begin{lstlisting}
let next_exam () =
  if Runtime.context.Solve([| Runtime.True |], null) then
    "no exam"
  else
    raise <| new InconsistentContext()
let next_exam () =
  let solution2 = new Dictionary<string, obj>()
  solution2.["exam"] <- new Variable()
  if Runtime.context.Solve([| physician_exam(phy, solution2.["exam"]);
                           patient_active_exam(pat, solution2.["exam"]) |],
                           solution2) then
    solution2.["exam"]
  else
    next_exam ()
printfn "%s can perform %s on %s" phy (next_exam ()) pat
\end{lstlisting}
Recall that parameters are evaluated in the context where they are referred to (line 16 in the original code), in a lazy way. 
This is rendered by the application \lstinline+next_exam ()+ in the \lstinline+printfn+ statement above.  
This resolves to the innermost definition of the function \lstinline+next_exam+, which checks the goal at lines 14-15. 
If the runtime finds a solution, the \lstinline+then+ branch evaluates to the goal variable \lstinline+ctx?exam+; otherwise  the task of determining the binding is delegated to the outermost \lstinline+let+. 
If even the outermost binding fails, an exception is raised to signal that the application is unable to adapt to the current context. 
However, here this will never be the case, because the predicate \lstinline+Runtime.True+ always holds.

\section{Discussion}
  % !TEX root = ../main.tex

Here, we briefly discuss the relevant literature on COP languages, and we refer the reader to the survey by Salvaneschi et al.~\cite{SalvaneschiGP13} on the design of languages, and to that by Appeltauer et al.~\cite{Appeltauer:2009} on some implementations.
Most of the proposals of COP languages (to cite just a few: \emph{JCop}~\cite{Appeltauer2013}, \emph{ContextL}~\cite{CostanzaH05},  \emph{Javanese}~\cite{Kamina:2013b}, \emph{Subjective-C}~\cite{Gonzales:2010}, \emph{PyContext}~\cite{vonLowis:2007}) describe the properties of the context as a stack of layers.
A layer can roughly be seen as an elementary proposition that  drives adaptation and that can be activated or deactivated at runtime.
Our context is instead a knowledge base, that offers primitives for easily storing and retrieving contextual data through Datalog queries.
Consequently, adaptation is driven on the basis of possibly complex deductions on the knowledge base.
We note that a (functional) language linked with a database system does not suffice for implementing adaptation to the context easily and directly, unless equipped with a deductive engine.
In the existing implementations, behavioural variations are often implemented as partially defined methods, and are not first-class (except for, e.g., \emph{ContextL}~\cite{CostanzaH05}), while ours are and it is well known that this feature improves code modularisation.
As a final remark,  many COP languages include a $\code{proceed}$ construct,  a sort of $\code{super}$ invocation in object oriented languages~\cite{HirschfeldCN08}, typically used for composing active behavioural variations.
This construct is strictly related to the idea of representing the context as a stack of layers, and it is unclear whether it makes sense to introduce a similar construct also in a full-fledged declarative context as ours.
Nevertheless, one could consider to implement a construct similar to \emph{call-next-method}~\cite{Vallejos:2010}, in order to run the next case with a satisfied goal, within the active behavioural variation.

%\medskip
Our implementation of \coda\ took advantage of F\#, both for the adaptation component and for the knowledge base.
The F\# compiler is stable and generates optimised bytecode; it is officially supported by Microsoft and fully integrated inside the .NET environment (and Mono, its open-source counterpart).  
F\# applications can readily run on all platforms (computers or mobile devices) supported by the many libraries and modules of CLR (or Mono). 

Our strategy has been to identify in the \coda\ code the constructs not native in F\# through suitable metaprogramming annotations. 
These annotations drive a JIT compilation of the adaptation constructs by reflecting over the code, while the rest of the code is compiled directly.
Therefore the whole compiler of \coda\ integrates the original F\# with the JIT compilation steps.
As a result, \coda\ becomes an ordinary .NET library, usable by any other (F\# compatible) application.
This prototypical implementation of \coda\ shows that the .NET type system allows us to solve the impedance mismatch between the functional and the logical components of \coda\ with a minimal effort.

Our compilation strategy is general and could be followed regardless
of the host implementation language, although some implementation
details, such as the generation of new identifiers, may require more
involved realisations in host languages other than F\#.  
In addition, our implementation of \coda-specific constructs would work independently from the ability to perform JIT compilation in the hosting framework.
Consequently, an Ahead-Of-Time (AOT) compilation approach would work as well, so possibly giving feedback to the programmer.
Typically, this would require to implement suitable static analyses, e.g.\ the two-step analysis of~\cite{SEFM14} that guarantees reliable adaptation of the application to contexts it will be hosted at runtime.
Both in JIT and AOT compilation, our approach allows for other different extensions, e.g.\
to enable code to easily interact with a model checker or other verifiers.

%\medskip

As a final remark, we recall that COP has been proposed by~\cite{Salvaneschi11} as a basis for implementing the software architecture of autonomic element proposed by~\cite{KephartC03}.
During an execution run, our JIT implementation only compiles those code fragments that the autonomic component needs to adapt, and skips those functions that are not invoked.

%%%%%%

\section{Conclusions}
  % !TEX root = ../main.tex

\label{sec:concl}

We presented an extension of F\# implementing the COP language \coda, that has a functional component for computing and a logical one for representing and querying the context.
Our implementation exploits metaprogramming mechanisms, such as reflection and quotation, to build a  JIT compiler.
Functions containing  \coda\ adaptation code are marked by the programmer, so driving an automatic translation to pure F\# code.
Our approach guarantees us a natural interaction between the code using the context and legacy code. 
This is particularly valuable since \coda\ code can run on all platforms supported by .NET and can access to all its libraries.
Besides the implementation of \coda\ itself, our JIT compilation schema scales to other host languages, and also to other different extensions.

We plan to extend our work along different lines.
First, we will equip our implementation with the two-step static analysis of~\cite{SEFM14}.
Besides the case studies available\,\footnote{ %
See  \url{https://github.com/vslab/fscoda}.}
we will assess our approach on other larger case studies.
Also benchmarking and comparing our implementation with others in the literature~\cite{GhezziACM12} is of interest.

More on the linguistic aspects, in the current setting the context is only updated by the applications, either by $\code{tell}$ or $\code{retract}$.
As experienced on the case study of the e-Healthcare system however, the context can evolve independently of the applications, emitting events to signal the changes --- implicitly representing the presence of many different applications sharing the same context.
The literature has a great deal of work on this topic, among which~\cite{Aotani:2011,engineer:2012}.
The major extensions to \coda\ for supporting these aspects include at least the ability of handling the concurrency between the context evolution and the running application, as well as primitives for react and adapt to events.

\bibliographystyle{eptcs}
\bibliography{nbiblio}

\end{document}